\documentclass{article}

\usepackage{arxiv}

\usepackage{amssymb}
\usepackage{amsmath}
\usepackage[utf8]{inputenc} 
\usepackage[T1]{fontenc}    
\usepackage{hyperref}       
\usepackage{url}            
\usepackage{booktabs}       
\usepackage{amsfonts}       
\usepackage{nicefrac}       
\usepackage{microtype}      
\usepackage{xcolor}         
\usepackage{algorithm}
\usepackage{algpseudocode}
\usepackage{graphicx}
\usepackage{multirow}
\usepackage[flushleft]{threeparttable}
\usepackage{float}
\usepackage{caption}
\usepackage{mathrsfs}

\title{Exp4Fuse: A Rank Fusion Framework for Enhanced Sparse Retrieval using Large Language Model-based Query Expansion}

\author{
 Lingyuan Liu \\
  City University of Hong Kong \\
  \texttt{ly.liu@my.cityu.edu.hk} \\
   \And
 Mengxiang Zhang\footnote{$*$} \\
  The University of Hong Kong \\
  \texttt{mxzhang6@connect.hku.hk} \\
}

\begin{document}
\maketitle
\begin{abstract}
Large Language Models (LLMs) have shown potential in generating hypothetical documents for query expansion, thereby enhancing information retrieval performance. However, the efficacy of this method is highly dependent on the quality of the generated documents, which often requires complex prompt strategies and the integration of advanced dense retrieval techniques. This can be both costly and computationally intensive. To mitigate these limitations, we explore the use of zero-shot LLM-based query expansion to improve sparse retrieval, particularly for learned sparse retrievers. We introduce a novel fusion ranking framework, Exp4Fuse, which enhances the performance of sparse retrievers through an indirect application of zero-shot LLM-based query expansion. Exp4Fuse operates by simultaneously considering two retrieval routes—one based on the original query and the other on the LLM-augmented query. It then generates two ranked lists using a sparse retriever and fuses them using a modified reciprocal rank fusion method. We conduct extensive evaluations of Exp4Fuse against leading LLM-based query expansion methods and advanced retrieval techniques on three MS MARCO-related datasets and seven low-resource datasets. Experimental results reveal that Exp4Fuse not only surpasses existing LLM-based query expansion methods in enhancing sparse retrievers but also, when combined with advanced sparse retrievers, achieves SOTA results on several benchmarks. This highlights the superior performance and effectiveness of Exp4Fuse in improving query expansion for sparse retrieval. 
\footnotetext{Corresponding author.}
\footnote{The code for our method is publicly available at \href{https://github.com/liuliuyuan6/Exp4Fuse}{https://github.com/liuliuyuan6/Exp4Fuse}.}
\end{abstract}

\section{Introduction}\label{sec:01}

Information retrieval is fundamental for extracting relevant documents from large databases and serves as a key component in various applications, including search engines, dialogue systems \cite{yuan2019multi}, question-answering platforms \cite{qu2020rocketqa, yang2023enhancing}, recommendation systems \cite{zhao2023recommender}, and retrieval-augmented generation \cite{zhang2022retgen}. The core objective of information retrieval is to index the documents within a collection and process user queries efficiently. Given a user's query, the system searches the index for documents that match the query terms and ranks these documents based on their relevance to the query. 

\begin{figure*}[t]
\centering
 \includegraphics[scale = 1.0]{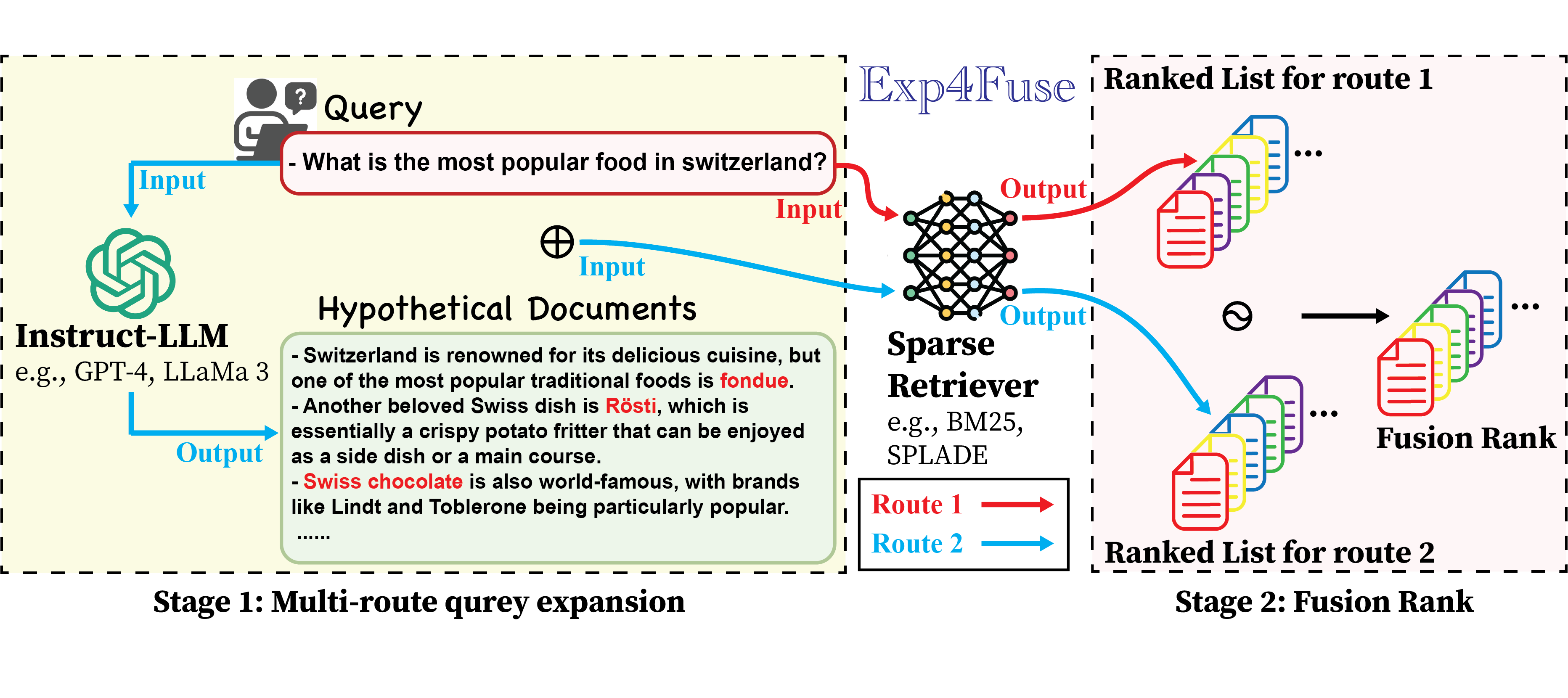} 
 \caption{\small An illustration of our Exp4Fuse framework. Exp4Fuse operates by simultaneously considering two retrieval routes—one based on the original query and the other on the LLM-augmented query. It then generates two ranked lists using a sparse retriever and fuses them using a modified reciprocal rank fusion method.}\label{fig:Exp4Fuse}
\end{figure*}

Query expansion (QE), a key technique refers to reformulate the original query with additional terms to bridge the gap between the user's query and the relevant documents \cite{abdul2004umass}, were widely used in enhancing performances of sparse and dense retrieval methods in information retrieval.
Initially, It was developed by using pseudo-relevance feedback or external knowledge sources.
However, its effectiveness is highly depend on the quality of the initial retrieval results.
With the emergence of large language models (LLMs) such as GPT-3 and LLaMA, significant progress has been made in generating fluent and realistic responses. Pre-trained on extensive corpora, LLMs excel in natural language understanding and generation.
Therefore, it inspired some studies in using LLMs for query expansion in sparse and dense retrieval methods, such as, HyDE \cite{gao2022precise}, query2doc \cite{wang2023query2doc} and LameR \cite{shen2024retrieval}. 
While these methods have shown empirical effectiveness, they also present certain limitations.

LLM-based QE face several limitations: i) outdated corpora in memory, as the LLM's training data may not reflect the most recent or up-to-date information, ii) generation of unreliable text, and iii) inability to specify the text domain. These issues impact the quality and credibility of pseudo-documents, affecting QE performance in dense and sparse retrieval. HyDE, a zero-shot LLM-based QE method, shows effective performance with dense retrievers but performs poorly with sparse retrievers. Conversely, query2doc, a few-shot LLM-based QE method, is effective with both sparse and dense retrieval methods. However, strong retrievers, such as learned sparse retrievers, may not benefit as much as weaker ones. \cite{jagerman2023queryexpansionpromptinglarge} demonstrated that the performance of prompting LLMs for QE is highly sensitive to the prompt shape. Complex prompt templates, such as Chain-of-Thought, exhibit the best performance. Similarly, the performance of LameR, which uses a question-answer prompt strategy, depends heavily on the quality of initial retrieval to formulate a high-quality prompt. These LLM-based QE methods often employ time-consuming strategies to improve the quality and credibility of pseudo-documents generated by LLMs, thereby enhancing QE performance in dense and sparse retrieval. However, this approach amplifies the inherent weaknesses of LLM-based QE, such as high costs and computational intensity, further limiting their practical deployment and efficiency, particularly when combined with dense retrievers.

From a practical perspective, it might be more suitable to combine LLM-based query expansion with sparse retrievers, which are lighter and faster, although their performance may not match that of dense retrievers. Recent studies have sought to transform traditional sparse retrievers into learned sparse retrievers using strategies such as distillation, hard-negative mining, and Pre-trained Language Model initialization \cite{formal2022distillation}. These learned sparse retrievers have shown superior or competitive performance compared to advanced dense retrievers on certain information retrieval benchmarks while remaining relatively lightweight and fast. Therefore, combining LLM-based query expansion with learned sparse retrievers could potentially yield competitive or even SOTA performance with lower time and computational costs. However, our observation revealed that traditional LLM-based query expansion methods, such as HyDE and query2doc, do not consistently enhance the performance of learned sparse retrievers. These methods may offer minor improvements under some metrics or fail to improve, and sometimes even degrade, the performance of advanced sparse retrievers, especially when used in a few-shot or zero-shot manner. Even with complex prompt strategies, LameR only achieves trivial improvements. Hence, the straightforward combination of LLMs with advanced sparse retrievers presents challenges, motivating the development of this work.

To further enhance sparse retrievers, particularly learned sparse retrievers, using LLM-based query expansion, we propose Exp4Fuse, a rank fusion framework (See Figure \ref{fig:Exp4Fuse}). This framework indirectly improves the performance of sparse retrievers through zero-shot LLM-based query expansion. It operates in two stages. In the first stage, there are two retrieval routes using a similar sparse retriever. The original route follows the traditional retrieval process, where the original query is input directly into the sparse retriever to rank relevant documents. The query expansion route involves zero-shot LLM-based query expansion, where the original query is input into LLMs to generate hypothetical documents. These documents are then used to augment the original query, which is subsequently input into the sparse retriever to rank relevant documents.
In the second stage, the document rankings from the two retrieval routes are fused using a modified reciprocal rank fusion method. The fused ranking is considered the final retrieval output of the Exp4Fuse framework.

We comprehensively evaluate the Exp4Fuse framework alongside basic and learned sparse retrievers. Under identical experimental conditions, we compare Exp4Fuse with existing LLM-based query expansion methods for various sparse retrievers, including query2doc and LameR, and for dense retrievers, including HyDE. Additionally, we compare it with the SOTA dense retriever and the SOTA multi-stage retrieval system - the retrieval \& rerank pipeline, across the MS MARCO dev dataset \cite{bajaj2016ms}, two TREC DL datasets \cite{craswell2020overview, DBLP:journals/corr/abs-2102-07662} for in-domain analysis, and seven low-resource datasets from the BEIR benchmark \cite{thakur2021beir} for out-of-domain analysis. These datasets encompass a range of tasks, including web search, question answering, and fact verification. Experimental results demonstrate that Exp4Fuse outperforms existing LLM-based query expansion methods for enhancing sparse retrievers, particularly for learned sparse retrievers, across most datasets and evaluation metrics. Furthermore, combining Exp4Fuse with advanced learned sparse retrievers outperforms some SOTA baselines and remains competitive with others, underscoring the high performance of Exp4Fuse.

Overall, the contributions can be summarized as follows:

\begin{itemize}
  \item  We propose Exp4Fuse, a query expansion method using a LLM to enhance sparse retrievers. Exp4Fuse fuses two sets of retrieved document ranks from the same sparse retriever: one based on the original query and the other on an LLM-based zero-shot query expansion, to generate final retrieved document ranks. This method benefits from indirect LLM-based QE and combines results from different query formats, yielding high-quality retrieval outcomes.

   \item  Exp4Fuse can effectively perform zero-shot QE for various sparse retrievers, particularly learned sparse retrievers. Extensive experiments  demonstrate that Exp4Fuse outperforms existing LLM-based query expansion methods. Furthermore, when combined with advanced sparse retrievers, Exp4Fuse surpasses some SOTA baselines and remains competitive with others.
\end{itemize}

\section{Related Work}\label{sec:02}

\subsection{Sparse Retrieval}\label{subsec:0201}
Sparse retrievers like BM25 \cite{robertson2009probabilistic} rank documents by matching terms in the query and document, considering term frequency and inverse document frequency. They are efficient, interpretable, and handle large vocabularies well but suffer from lexical mismatch issues. To address this, methods like query and document expansion \cite{nogueira2019document}, such as docT5query \cite{nogueira2019doc2query}. Learned sparse retrievers like uniCOIL \cite{lin2021few}, SPLADE \cite{formal2021splade}, and SLIM \cite{li2023slim} improve retrieval by contextualizing term representations and incorporating sparse activations. Advanced models like SPLADEv2 \cite{formal2021splade} use techniques like distillation and hard negative mining for SOTA results. However, improving these models with LLMs directly is challenging. Our Exp4Fuse framework offers a simple, flexible solution by using a single sparse retriever to rank documents based on both original and LLM-augmented queries, avoiding dense retrievers' high memory and time costs. Exp4Fuse is compatible with various sparse retrievers, especially learned ones.

\subsection{LLM-based Query Expansion}\label{subsec:0202}
LLM-based query expansion leverages LLM to generate pseudo-references or potential answers, thereby enhancing queries for improved retrieval. The core concept involves using LLMs in a zero-shot, few-shot, or complex prompting manner to create hypothetical documents, which are then concatenated with the original query for use in retrieval tasks.
For instance, HyDE \cite{gao2022precise} employs LLMs for zero-shot query expansion to boost dense retrieval, while query2doc \cite{wang2023query2doc} uses few-shot query expansion to improve both sparse and dense retrieval. 
Research by \cite{jagerman2023queryexpansionpromptinglarge} explored various prompt strategies for LLM-based query expansion in sparse retrieval, and LameR \cite{shen2024retrieval} utilized complex question-answer prompts for enhancing both retrieval types.
Despite the improvements in dense and sparse retrievers, existing LLM-based query expansion methods face challenges, such as limited gains for strong retrievers (e.g., learned sparse retrievers and fine-tuned dense retrievers) and high computational and time costs.
Our proposed Exp4Fuse addresses these issues by employing LLMs for zero-shot query expansion in a cost-effective manner.

\subsection{Fusion Retrieval}\label{subsec:0203}
Extensive research in information retrieval reveals that dense retrievers excel at modeling semantic similarity but may struggle with exact matches and long documents, where sparse retrievers are more effective. Recent studies have attempted to fuse dense and sparse retrievers, combining their strengths. Fusion models typically merge results using a convex combination of lexical and semantic scores \cite{ma2020hybrid} or the reciprocal rank fusion method \cite{cormack2009reciprocal}. For example, \cite{chen2022out} developed a fusion framework using reciprocal rank fusion, combining neural passage retrieval \cite{lu2021multi} with BM25 variants.
Most efforts focus on multi-stage retrieval systems, which involve an initial retrieval stage followed by several re-ranking stages. Sparse retrievers like BM25 efficiently generate initial candidate sets, while dense models re-rank the most promising candidates, enhancing recall and ranking quality \cite{nogueira2019multi}. 
However, few studies explore fusion retrieval using a single retriever type, such as RepBERT \cite{zhan2020repbert}. While existing fusion methods achieve high performance, they are often costly and computationally intensive due to multiple retrieval models and reliance on dense models, especially when using LLMs to enhance fusion. In contrast, by employing LLMs for zero-shot query expansion and using only a single sparse retriever, Exp4Fuse requires lower computational and memory resources, making it easy to deploy.

\section{Methodology}\label{sec:03}
In this section, we detail the Exp4Fuse framework, illustrated in Figure \ref{fig:Exp4Fuse}. The framework consists of two stages: multi-route query expansion and fusion ranking. In the first stage, we employ two retrieval routes using a sparse retriever. The original route involves the traditional retrieval process, where the original query is input into the sparse retriever to generate a ranked list of documents ($I_{oq}$). The LLM-based QE route involves inputting the original query into an LLM to generate a hypothetical document, which is then concatenated with the original query and input into the sparse retriever to produce another ranked list ($I_{eq}$). In the second stage, the two ranked document lists from the different routes are fused using a modified reciprocal rank fusion method to generate the final ranked list. Details on the zero-shot LLM-based QE are provided in section \ref{sec:0301}, and the fusion ranking method is discussed in section \ref{sec:0302}.

\subsection{LLM-based query expansion}\label{sec:0301}
In the zero-shot LLM-based QE route of the Exp4Fuse framework, user queries are augmented using a straightforward approach. Upon receiving a query $q_{o}$, Exp4Fuse applies a simple zero-shot prompt to generate a hypothetical document, denoted as $r_{q}$, which is then concatenated with the original query as input for the subsequent sparse retriever.
Sparse retrievers typically evaluate relevance by analyzing lexical overlaps, making them sensitive to word frequency. The hypothetical document generated by the LLM is generally longer than the original query. A simple combination of the original query and the generated document might not be effective for sparse retrieval because it could imbalance the influence of each element in the augmented query.
To address this issue, we implement a weighting adjustment strategy that increases the length of the original query by repeating it. This balances the influence of the original query and the hypothetical document. The adjustment is governed by a weight $\lambda$. We enhance the query by repeating the original query $\lambda$ times and then concatenating it with the hypothetical document.

\begin{equation}\label{01}
q_{e} = \textup{concat} (q_{o} \times \lambda, r_{q}).
\end{equation}

\subsection{Fusion Rank}\label{sec:0302}
In the second stage of the Exp4Fuse framework, the fusion rank aims to combine two ranked lists of retrieved documents ($I_{eq}$ and $I_{oq}$), resulting in the final ranked list $I_{fq}$. This is achieved using an improved reciprocal rank fusion method \cite{cormack2009reciprocal}, which calculates and ranks the scores for each document based on their positions in the two lists.
Our choice of the reciprocal rank fusion method is motivated by the principle that while highly-ranked documents are crucial, lower-ranked documents still hold significance, unlike in exponential ranking functions. We enhance the existing method by incorporating an adaptive weight strategy that adjusts the final rank score of a document based on its presence in both lists and the relative importance of the two retrieval methods. This adjustment ensures that documents retrieved by both routes are more likely to be included in the final ranked list.
The improved reciprocal rank fusion method uses the following scoring formula:

\begin{equation}\label{01}
FR_{score} = (w_{i} + \frac{n}{10} ) \cdot \sum_{i=1}^{2} \frac{1}{k+r_{i}},
\end{equation}
where $k=60$ is a constant fixed during a pilot study to mitigate the impact of outlier rankings. $r_{i}$ represents the rank of the document in the retrieval list $i$ ($i=1$ for $I_{oq}$ and $i=2$ for $I_{eq}$). $w_{i}$ represents the weight for the retrieval list $i$. $n$ indicates the number of times the document appears in the two lists, with $n \in \{1, 2 \}$.

\section{Experiments}\label{sec:04}
In this section, we conduct comprehensive experimental evaluations using Exp4Fuse and compare it with existing mainstream competitors to demonstrate the proposed method's effectiveness and efficiency.

\subsection{Experimental Setup}\label{subsec:0401}

\paragraph{Datasets}
Following the protocols of existing LLM-based QE methods \cite{wang2023query2doc, shen2024retrieval}, we evaluate our method on two types of datasets pertinent to information retrieval tasks. The first type includes in-domain datasets: MS-MARCO dev \cite{bajaj2016ms} and its sub-datasets, TREC-DL-2019 \cite{craswell2020overview} and TREC-DL-2020 \cite{DBLP:journals/corr/abs-2102-07662}. The second type consists of out-of-distribution datasets, comprising a diverse collection of seven low-resource datasets from the BEIR benchmark \cite{thakur2021beir}, including DBPedia, FiQA, News, NQ, Robust04, Touche2020, and Scifact. Distinct instructions are utilized for each dataset, maintaining a consistent structure but varying quantifiers to control the form of the generated hypothetical documents. Detailed instructions can be found in Appendix \ref{A:0103}.

\paragraph{Implementation Details}
We employ GPT4-mini \cite{achiam2023gpt} as the backbone model to generate hypothetical documents for QE. These documents are sampled with a temperature of 0.6, top-p of 0.9, and a maximum of 128 tokens for open-ended generation. We select four types of sparse retrievers and their variants to examine the performance of the Exp4Fuse framework, including BM25, uniCOIL, SPLADEv2, and SLIM. For all searches, we use the Pyserini toolkit \cite{lin2021pyserini} with default settings, retrieving the top 1000 documents as ranked lists for subsequent fusion. In our experiments, unless specified otherwise, Exp4Fuse uses $\lambda = 5$ for LLM-based query expansion and sets $w_{1} = w_{2} = 1$ for the fusion rank score calculation. All experiments are conducted on an NVIDIA L20 GPU with 48GB of memory.

\paragraph{Baselines and Competitors}
We compare Exp4Fuse with two types of baseline approaches to demonstrate its effectiveness:

\textit{Basic sparse retriever}: This approach includes basic sparse retrievers and simple variants without any learning strategy. Specifically, BM25 \cite{robertson2009probabilistic} and BM25 + docT5query \cite{nogueira2019doc2query}, where retrieved documents are expanded using docT5query, then indexed and ranked by BM25.

\textit{Learned sparse retriever}: This approach includes learned sparse retrievers derived from associated sparse retrievers using different training strategies such as distillation, hard negative mining, pre-training, or combinations thereof. Specifically:
uniCOIL \cite{lin2021few}, trained with BM25 hard negatives from MS MARCO Passages,
SLIM$^{++}$ \cite{li2023slim}, trained with cross-encoder distillation \cite{hofstatter2020improving} and hard-negative mining,
SPLADE$^{++}$-v1 \cite{formal2022distillation}, initialized from a pre-trained CoCondenser \cite{gao2021unsupervised} checkpoint and trained with ensemble-mining and distillation,
SPLADE$^{++}$-v2 \cite{formal2022distillation}, initialized from a pre-trained CoCondenser checkpoint and trained with self-mining and distillation.

Secondly, we compare Exp4Fuse with three types of competitor approaches to demonstrate its performance:

\textit{LLM-augmented sparse retriever}: This approach includes existing LLM-based QE methods to enhance basic sparse retrievers, such as BM25 + query2doc \cite{wang2023query2doc} and BM25 + LameR \cite{shen2024retrieval}.

\textit{Advanced dense retrievers and their LLM-augmented variants}: This approach involves high-performing and SOTA dense retrievers with fully-supervised training and their variants using existing LLM-based QE methods, including TAS-B \cite{hofstatter2021efficiently}, coCondenser \cite{gao2021unsupervised}, fine-tuned Contriever + HyDE \cite{gao2022precise}, SimLM \cite{wang2022simlm}, SimLM + query2doc, and SimLM + LameR.

\textit{Advanced multi-stage retrieval systems}: This approach considers current mainstream multi-stage retrieval systems — the retrieval \& re-rank pipelines— that have demonstrated high and SOTA performance in information retrieval tasks. Examples include RepLLaMA (retriever) + RankLLaMA (re-ranker) \cite{ma2024fine}, which fine-tune the latest LLaMA model both as a dense retriever and re-ranker using the MS MARCO datasets, monoT5-3B (retriever) + BM25 (re-ranker) \cite{nogueira2020document}, uniCOIL (retriever) + ColBERTv2/CQ \cite{yang2022compact} (re-ranker), and SPLADE (retriever) + ColBERT/BKL (re-ranker) trained with balanced KL divergence \cite{yang2023balanced}.

\paragraph{Metrics}
Similar to prior research \cite{wang2022simlm, wang2023query2doc}, we use the following evaluation metrics: $MAP$, $nDCG@10$, and $Recall@1k$ for TREC DL 2019 and 2020, $MRR@10$ and $Recall@1k$ for MS-MARCO for in-domain analysis, and $nDCG@10$ for the BEIR datasets for out-of-distribution evaluation.

\subsection{Results}\label{subsec:0402}

\begin{table*}[h]\scriptsize
    \centering
    \begin{tabular}{l|ll|lll|lll} \hline
        ~ & \multicolumn{2}{c}{\textbf{MS MARCO dev}} & \multicolumn{3}{c}{\textbf{TREC DL 19}}   & \multicolumn{3}{c}{\textbf{TREC DL 20}} \\ 
        ~ & MRR@10 & R@1k  &  MAP & nDCG@10 &  R@1k & MAP & nDCG@10 &  R@1k \\ \hline
\multicolumn{9}{l}{\textit{Basic sparse retriever and the associated LLM-augmented variant} }  \\ 
BM25 \cite{robertson2009probabilistic} & 18.4$^{\ast}$ & 85.7$^{\ast}$  & 30.1$^{\ast}$ & 50.6$^{\ast}$ & 75.0$^{\ast}$ & 28.6$^{\ast}$ & 48.0$^{\ast}$ & 78.6$^{\ast}$ \\ 
~~~~~~+\textbf{Exp4Fuse} &  20.7$^{+2.3}$ & 91.3$^{+5.6}$ & 38.7$^{+8.6}$ & 62.0$^{+12.4}$ & 87.0$^{+12.0}$ & 36.3$^{+12.3}$ & 56.6$^{+12.6}$ & 87.3$^{+8.7}$ \\ 
docT5query \cite{nogueira2019doc2query} & 27.2$^{\ast}$ & 94.7$^{\ast}$ & 40.3$^{\ast}$ & 64.2$^{\ast}$ & 83.1$^{\ast}$ & 40.7$^{\ast}$ & 61.9$^{\ast}$ & 84.5$^{\ast}$ \\ 
~~~~~~+\textbf{Exp4Fuse}  & \textbf{28.7$^{+1.5}$} & \textbf{96.4$^{+1.7}$} & \textbf{47.5$^{+7.2}$} & \textbf{68.5$^{+4.3}$} & \textbf{87.8$^{+4.7}$} & \textbf{45.7$^{+5.0}$} & \textbf{64.2$^{+2.3}$} & \textbf{89.3$^{+4.8}$} \\ \hline
\multicolumn{9}{l}{\textit{Learned sparse retriever }}  \\                   
uniCOIL \cite{lin2021few} & 35.0$^{\ast}$ & 95.8$^{\ast}$ & 46.1$^{\ast}$ & 70.2$^{\ast}$ & 82.9$^{\ast}$ & 44.3$^{\ast}$ & 67.5$^{\ast}$ & 84.3$^{\ast}$ \\ 
~~~~~~+\textbf{Exp4Fuse}  & 36.4$^{+1.4}$  & 97.4$^{+1.6}$  & 50.1$^{+4.0}$ & 73.6$^{+3.4}$ & 86.5$^{+3.6}$ & 47.2$^{+2.9}$ & 70.9$^{+3.4}$ & 87.7$^{+3.4}$ \\ 
SLIM$^{++}$ \cite{li2023slim}  & - & - & 48.3$^{\ast}$ & 72.5$^{\ast}$ & 86.8$^{\ast}$ & 48.7$^{\ast}$ & 69.2$^{\ast}$ & 87.1$^{\ast}$ \\ 
~~~~~~+\textbf{Exp4Fuse} & - & - & 50.6$^{+2.3}$ & 76.6$^{+4.1}$ & 87.2$^{+0.4}$ & 48.8$^{+0.1}$ & 70.4$^{+1.2}$ & 88.6$^{+1.5}$ \\ 
SPLADE$^{++}$-v1 \cite{formal2022distillation} & 36.9$^{\ast}$ & 97.9$^{\ast}$ & 50.5$^{\ast}$ & 73.1$^{\ast}$ & 87.3$^{\ast}$ & 50.0$^{\ast}$ & 72.0$^{\ast}$ & 90.0$^{\ast}$ \\ 
~~~~~~+\textbf{Exp4Fuse} & \textbf{39.8$^{+2.9}$} & 98.6$^{+0.7}$ &  \textbf{56.7$^{+6.2}$} & \textbf{77.6$^{+4.5}$} & 93.3$^{+6.0}$ & 54.3$^{+4.3}$ & 73.3$^{+1.3}$ & 92.9$^{+2.9}$ \\ 
SPLADE$^{++}$-v2 \cite{formal2022distillation} & 36.8$^{\ast}$ & 98.0$^{\ast}$ & 50.0$^{\ast}$ & 73.6$^{\ast}$ & 87.6$^{\ast}$ & 51.4$^{\ast}$ & 72.8$^{\ast}$ & 90.2$^{\ast}$ \\ 
~~~~~~+\textbf{Exp4Fuse} & 39.6$^{+2.8}$ & \textbf{98.9$^{+0.9}$} & 56.6$^{+6.6}$ & 77.1$^{+3.5}$ & \textbf{93.9$^{+6.3}$} & \textbf{55.8$^{+4.4}$} &  \textbf{73.8$^{+1.0}$} & \textbf{93.8$^{+3.6}$} \\ \hline
\multicolumn{9} {l}{\textit{Advanced dense retriever and the associated LLM-augmented variant} }  \\
TAS-B \cite{hofstatter2021efficiently}   &  34.0 & 97.5  & - & 71.2 & 84.3 & - & 69.3 & - \\ 
coCondenser \cite{gao2021unsupervised} &  38.2 & 98.4 & - & 71.7 & 82.0 & - & 68.4  & 83.9  \\ 
Contriever$^{FT}$ + HyDE \cite{gao2022precise} & - & - & - & 67.4 & - & - & 63.5 & - \\ 
SimLM \cite{wang2022simlm} & 41.1 & 98.7  & - & 71.4 & - & - & 69.7 & - \\
~~~~~~+query2doc \cite{wang2023query2doc} & 41.5 & 98.8 & - & 72.9 & - & - & 71.6 & - \\ 
~~~~~~+LameR \cite{shen2024retrieval} & - & - & 54.9 & 76.5 & 91.1 & 55.7 & 75.8 & 89.5 \\ \hline
\multicolumn{9}{l}{\textit{Advanced multi-stage retrieval system} }  \\ 
monoT5-3B + BM25 \cite{zhang2024exploringbestpracticesquery} & - & - & - & 71.8  & - & - & 68.9  & - \\ 
RepLLaMA + RankLLaMA-13B \cite{ma2024fine} & \textbf{45.2} & \textbf{99.4} & - & 76.0  & - & - &  \textbf{77.9} & - \\ 
        uniCOIL + ColBERTv2/CQ \cite{yang2023balanced} & 38.7 & 95.8 & - & 74.6 & - & - & 72.6 & - \\ 
        SPLADE + ColBERT/BKL \cite{yang2023balanced} & 40.7 & 98.2 & - & 71.6 & - & - & 73.6 & - \\ \hline
    \end{tabular}
\caption{Results for web search on MS MARCO databases - MS MARCO dev and TREC DL 19/20. Best performing systems are marked \textbf{bold}. Results with $^{\ast}$ are from our reproduction with public checkpoints. }
\label{tab1}
\end{table*}

\paragraph{In-Domain evaluations} 
As presented in Table \ref{tab1}, our principal findings from the retrieval evaluations for in-domain datasets can be summarized as follows:

\textbf{1) Exp4Fuse enhances the performance of both basic and learned sparse retrievers.} Combining the Exp4Fuse framework with any sparse retriever, whether basic or learned, improves performance on three web search benchmark datasets—MS MARCO dev and TREC DL 2019/2020—across all metrics. This demonstrates the robustness and reliability of Exp4Fuse in enhancing sparse retrieval.

\textbf{2) Exp4Fuse outperforms other LLM-based QE methods in web search tasks.} Compared to other LLM-based QE methods in the context of basic sparse retrieval, Exp4Fuse combined with docT5query outperforms approaches like query2doc and LameR across all metrics on all web search benchmark datasets. In advanced retrieval, Exp4Fuse combined with SPLADE$^{++}$-v1 and SPLADE$^{++}$-v2 surpasses other LLM-based QE methods, when paired with strong dense retrievers like SimLM, across most metrics on three web search benchmarks. The only exceptions are a slight underperformance in MAP@10 on MS MARCO dev for query2doc and in nDCG@10 on TREC DL 2020. This underscores Exp4Fuse's effectiveness in enhancing sparse retrieval performance through LLM-augmented query expansion.

\textbf{3) Exp4Fuse combined with learned sparse retrievers matches strong baselines in web search tasks.} Compared to advanced retrieval methods, Exp4Fuse combined with SPLADE$^{++}$-v1 and SPLADE$^{++}$-v2 closely matches the performance of advanced and even SOTA methods. Notably, SPLADE$^{++}$-v1 + Exp4Fuse achieves SOTA performance on the TREC DL 2019 dataset across all metrics. Additionally, for other benchmark datasets, Exp4Fuse combined with SPLADE$^{++}$-v1 or SPLADE$^{++}$-v2 remains competitive with other SOTA methods, showcasing Exp4Fuse's superior effectiveness in enhancing sparse retrieval performance through LLM-based QE.

\paragraph{Out-of-Domain evaluations}

\begin{table*}[h]\scriptsize
    \centering
    \begin{tabular}{llllllll}
    \hline
        ~ & \textbf{DBPedia} & \textbf{FiQA} & \textbf{News} & \textbf{NQ} & \textbf{Robust04} & \textbf{Touche2020} & \textbf{Scifact} \\ 
        ~ & \multicolumn{7}{c}{nDCG@10}   \\ \hline
\multicolumn{8}{l}{\textit{Basic sparse retriever and the associated LLM-augmented variant} } \\ 
BM25 \cite{robertson2009probabilistic} & 31.8$^{\ast}$ & 23.6$^{\ast}$ & 39.5$^{\ast}$ & 30.6$^{\ast}$ & 40.7$^{\ast}$ & 44.2$^{\ast}$ & 67.9$^{\ast}$ \\ 
~~~~~~+\textbf{Exp4Fuse} & 36.1$^{+4.3}$ & 24.7$^{+1.1}$ & 44.8$^{+5.3}$ & 39.1$^{+8.5}$ & 46.5$^{+5.8}$ & \textbf{51.2$^{+7.0}$} & 68.8$^{+0.9}$ \\ 
docT5query \cite{nogueira2019doc2query} & 33.1$^{\ast}$ & 25.2$^{\ast}$ & 42.0$^{\ast}$ & 38.1$^{\ast}$ & 43.7$^{\ast}$ & 34.7$^{\ast}$ & 67.5$^{\ast}$ \\ 
~~~~~~+\textbf{Exp4Fuse}  & \textbf{38.9$^{+5.8}$} & \textbf{26.3$^{+1.1}$} & 48.7$^{+6.7}$ & \textbf{42.8$^{+4.7}$} & \textbf{46.9$^{+3.2}$} & 39.9$^{+5.2}$ & 71.3$^{+3.8}$ \\ \hline
\multicolumn{8}{l}{\textit{Learned sparse retriever }}  \\ 
uniCOIL \cite{lin2021few} & 33.8 & 28.9 & - & 42.5 & - & 29.8 & 68.6 \\ 
SPLADE$^{++}$-v1 \cite{formal2022distillation} & 43.7$^{\ast}$ & 34.7$^{\ast}$ & 41.7$^{\ast}$ & 53.7$^{\ast}$ & 46.6$^{\ast}$ & 24.6$^{\ast}$ & 70.4$^{\ast}$ \\ 
~~~~~~+\textbf{Exp4Fuse} & \textbf{47.2$^{+3.5}$} & \textbf{36.5$^{+1.8}$} & \textbf{42.5$^{+0.8}$} & \textbf{61.3$^{+7.6}$} & \textbf{53.1$^{+6.5}$} & \textbf{33.3$^{+8.7}$} & \textbf{73.8$^{+3.4}$} \\ \hline
\multicolumn{8}{l}{\textit{Advanced dense retriever and the associated LLM-augmented variant} }  \\ 
TAS-B \cite{hofstatter2021efficiently} & 38.4 & 29.6 & - & 46.5 & - & 22.2 & 64.4 \\ 
Contriever + HyDE \cite{gao2022precise} & 36.8 & 27.3 & - & - & - & - & 69.1 \\ 
SimLM \cite{wang2022simlm}  & 34.9 & - & - & - & - & 18.9 & 62.4 \\ 
~~~~~~+query2doc \cite{wang2023query2doc} & 38.3 & - & - & - & - & 25.6 & 59.5 \\ \hline
\multicolumn{8}{l}{\textit{Advanced multi-stage retrieval system} }  \\ 
monoT5-3B + BM25 \cite{zhang2024exploringbestpracticesquery} & 44.5 & - & 48.5 & - & \textbf{56.7} & 32.4 & \textbf{76.6} \\ 
RepLLaMA + RankLLaMA-13B \cite{ma2024fine} & \textbf{48.7} & \textbf{48.1} & - & \textbf{66.7} & - & 40.6 & 73.0 \\ \hline
    \end{tabular}
\caption{Out-of-domain results on 7 low-resource datasets from the BEIR benchmark. Best performing systems are marked \textbf{bold}. Results with $^{\ast}$ are from our reproduction with public checkpoints. }
\label{tab2}
\end{table*}

As presented in Table \ref{tab2}, our principal findings from the retrieval evaluations for out-of-domain datasets can be summarized as follows:

\textbf{1) Exp4Fuse enhances the performance of basic and learned sparse retrievers for low-resource retrieval.}
Exp4Fuse shows substantial improvements on the BEIR dataset, where queries are typically short and ambiguous. Regardless of the type of sparse retriever, combining the Exp4Fuse framework with sparse retrievers outperforms the baselines on seven BEIR databases. The degree of enhancement varies, being more pronounced for some datasets like Touche 2020 and less for others like FiQA.

\textbf{2) Exp4Fuse is competitive with strong baselines for low-resource retrieval.}
For basic sparse retrievers, Exp4Fuse + docT5query outperforms other QE strategies across most test datasets, except for News and Scifact. Notably, Exp4Fuse + BM25 achieves SOTA performance for low-resource retrieval on the Touche 2020 benchmark, measured by nDCG@10. For advanced retrievers, the combination of Exp4Fuse with SPLADE$^{++}$-v1 remains competitive with other SOTA methods.

\section{Analysis}\label{sec:05}

\paragraph{Generalizability} 

We conducted additional experiments using the open-source LLaMA3-8B-Instruct model. The results on MS MARCO dev, DL19, and DL20 are presented in Table \ref{tab3}, demonstrating the generality of our method.

\begin{table*}[h]\scriptsize
    \centering
\begin{tabular}{@{}l|ll|lll|lll@{}}
\toprule
\multicolumn{1}{c|}{\multirow{2}{*}{\textbf{Model}}} & \multicolumn{2}{c|}{\textbf{MS MARCO dev}}             & \multicolumn{3}{c|}{\textbf{TREC DL 19}}                                          & \multicolumn{3}{c}{\textbf{TREC DL 20}}                                          \\
\multicolumn{1}{c|}{}                                & \multicolumn{1}{c}{MRR@10} & \multicolumn{1}{c|}{R@1k} & \multicolumn{1}{c}{MAP} & \multicolumn{1}{c}{nDCG@10} & \multicolumn{1}{c|}{R@1K} & \multicolumn{1}{c}{MAP} & \multicolumn{1}{c}{nDCG@10} & \multicolumn{1}{c}{R@1K} \\ \midrule
BM25                                                 & 18.4                       & 85.7                      & 30.1                    & 50.6                        & 75.0                        & 28.6                    & 48.0                          & 78.6                     \\
+Exp4Fuse\_LLaMa                                     & 18.9                       & 90.6                      & 34.3                    & 59.7                        & 76.7                      & 31.5                    & 49.1                        & 82.6                     \\
uniCOIL                                              & 35.0                         & 95.8                      & 46.1                    & 70.2                        & 82.9                      & 44.3                    & 67.5                        & 84.3                     \\
+Exp4Fuse\_LLaMa                                     & 36.5                       & 96.5                      & 47.6                    & 73.6                        & 85.0                        & 46.1                      & 69.2                        & 84.8                     \\
SPLADE\_v1                                           & 36.9                       & 97.9                      & 50.5                    & 73.1                        & 87.3                      & 50.1                      & 72.0                          & 90.1                       \\
+Exp4Fuse\_LLaMa                                     & 38.6                       & 97.8                      & 51.3                    & 75.8                        & 92.1                      & 52.2                    & 73.1                        & 92.3                     \\
SPLADE\_v2                                           & 36.8                       & 98.1                        & 50.1                      & 73.6                        & 87.6                      & 51.4                    & 72.8                        & 90.2                     \\
+Exp4Fuse\_LLaMa                                     & 38.7                       & 98.8                      & 50.5                    & 75.4                        & 92.9                      & 53.4                    & 78.7                        & 93.8                     \\ \bottomrule
\end{tabular}
\caption{Results for Exp4Fuse using LLaMA3-8B-Instruct.}
\label{tab3}
\end{table*}

\paragraph{Ablation Study} To better understand the utility of Exp4Fuse, we conduct various experiments on the TREC DL 2019/2020 datasets to analyze the impact and effectiveness of each component within the architecture. The Exp4Fuse settings are described in Section \ref{subsec:0401}.

\begin{figure}[t]
\centering
 \includegraphics[scale = 0.12]{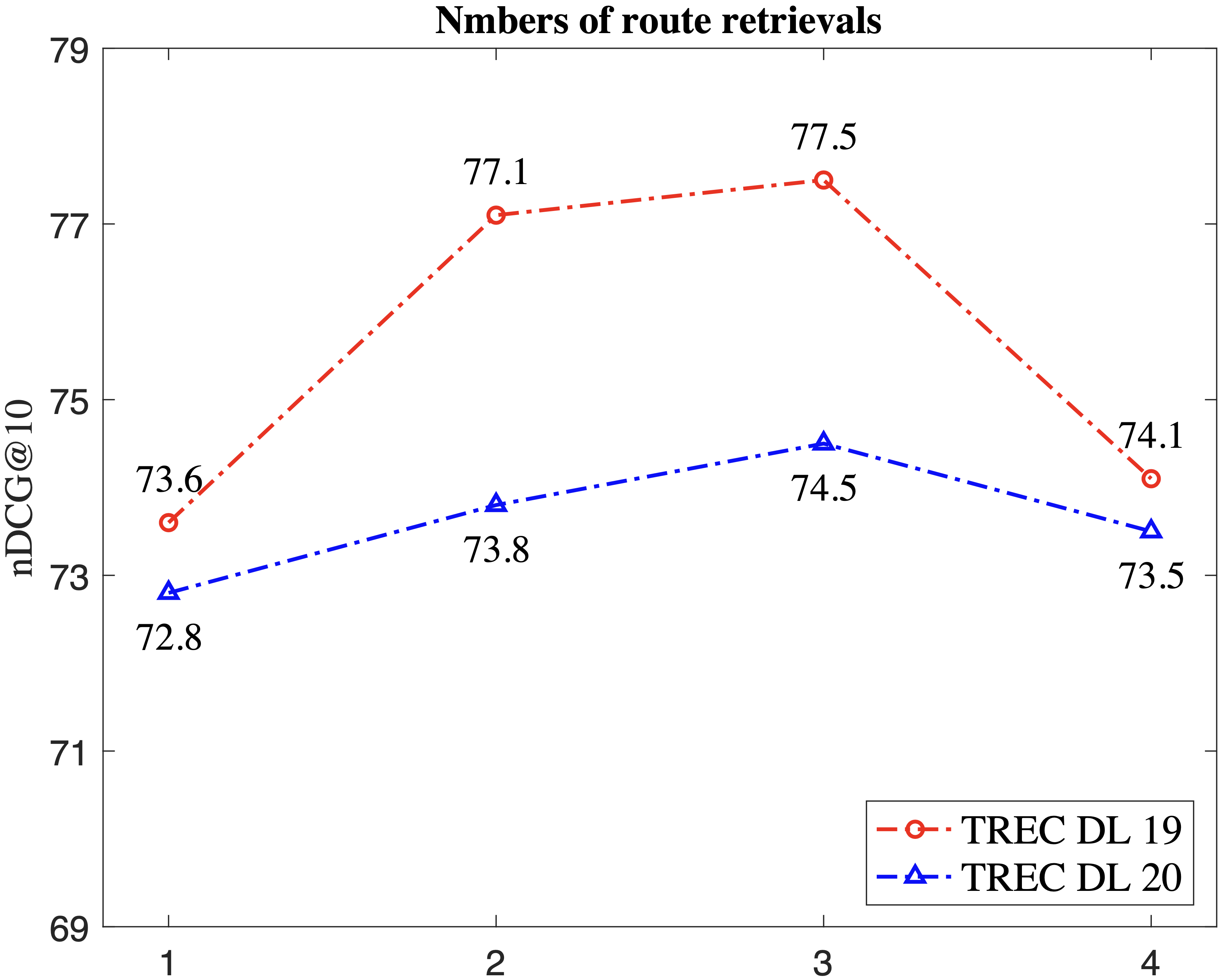} 
 \caption{Impact of the numbers of route retrievals.}\label{fig:nrr}
\end{figure}

\paragraph{1) Impact of the Number of Route Retrievals} In this experiment, we investigate the impact of the number of route retrievals on the performance of enhancing sparse retrievers by adding two variants of QE with zero-shot LLMs: multiple query expansion and step-back query expansion. Multiple query expansion, inspired by \cite{belkin1995combining}, augments the original query with different versions generated by LLMs. Step-back query expansion, inspired by \cite{zheng2023take}, augments the original query with high-level concepts and first principles generated by LLMs using step-back prompting. Detailed settings for these variants are provided in the Appendix \ref{A:0101} and \ref{A:0102}.

We consider four route retrievals based on different query inputs: the original query (OQ), hypothetical document query expansion (HDQ), multiple query expansion (MQ), and step-back query expansion (SBQ), using SPLADE$^{++}$-v2 as the backbone model to retrieve on the TREC DL 2019/2020 datasets. The Exp4Fuse framework is used to fuse the ranking lists from different numbers of routes: one (OQ), two (OQ + HDQ), three (OQ + HDQ + MQ), and four (OQ + HDQ + MQ + SBQ).

From Figure \ref{fig:nrr}, we can draw the following conclusions: a) In the Exp4Fuse framework, the performance of sparse retrievers on the TREC DL 2019/2020 datasets improves with an increase in the number of route retrievals from 1 to 3. However, adding more routes beyond 3 with LLM-based query expansion may decrease performance. A plausible explanation is that LLM-based QE can generate both relevant and irrelevant passages. Relevant passages benefit sparse retrievers by improving query-document matching, while irrelevant passages can degrade performance. Different route retrievals with LLM-based query expansion contain varying ratios of relevant and irrelevant passages. When the number of routes is low, relevant passages dominate and enhance performance. However, there is an upper limit to the number of relevant passages generated by LLMs. As the number of route retrievals increases, the impact of irrelevant passages becomes more significant, reducing the overall performance. b) There is a diminishing return in performance gains with each additional LLM-based route retrieval. This suggests that the combination of OQ and HDQ routes is cost-effective, balancing performance improvements with computational and time costs.

\paragraph{2) Necessity of Fusion Ranking} In this experiment, we evaluate the performance of each route retrieval in the Exp4Fuse framework to investigate the necessity of the fusion ranking stage. We consider the original query (OQ), hypothetical document query expansion (HDQ), and multiple query expansion (MQ) as inputs for sparse retrieval. SPLADE$^{++}$-v1 and SPLADE$^{++}$-v2 serve as backbone models for retrieval on the TREC DL 2019/2020 datasets. Additionally, we include benchmarks for the fusion results of OQ and HDQ, and OQ, HDQ, and MQ using the Exp4Fuse framework.

\begin{table}[!ht]\footnotesize
    \centering
    \begin{tabular}{lcccc}
    \hline
        ~ & \multicolumn{2}{c}{DL 19} & \multicolumn{2}{c}{DL 20} \\ 
        nDCG@10 & v1 & v2 & v1 & v2 \\ \hline
        original query & 73.1 & 73.6 & 72.0 & 72.8 \\ 
        hypothetical query & 67.8 & 66.9 & 65.0 & 64.4 \\ 
        multiple query & 70.5 & 71.9 & 69.1 & 69.5 \\ 
        original +hypothetical query & 77.6 & 77.1 & 77.1 & 73.8 \\ 
        OQ+HDQ+MQ & 77.7 & 77.8 & 74.5 & 74.5 \\ \hline
    \end{tabular}
    \caption{Necessity of fusion ranking on TREC DL 19/20 dataset. v1 represents SPLADE$^{++}$-v1, and v2 represents SPLADE$^{++}$-v2.}
    \label{tab3}
\end{table}

Table \ref{tab3} presents the performance of each route retrieval within the Exp4Fuse framework. As shown, using zero-shot LLM-based QE directly to enhance sparse retrieval may not always be effective and can sometimes negatively impact performance. This is likely because learned sparse retrieval models, trained on the MS MARCO dataset using original queries and documents, excel at matching these inputs for web search. Thus, the brute-force combination of the original query with LLM-generated passages may cause mismatches between queries and documents in learned sparse retrieval. These findings highlight the limitations of directly using zero-shot LLM-based QE to enhance sparse retrieval performance. They indicate the necessity of fusing the original query route with LLM-based QE routes for further improvement in learned sparse retrieval.

\section{Conclusion }\label{sec:06}
In this paper, we introduce the Exp4Fuse framework to enhance sparse retrieval through zero-shot LLM-based QE. Unlike existing methods, Exp4Fuse leverages LLM-generated knowledge indirectly for QE. Specifically, it considers two route retrievals: the original query route and the zero-shot LLM-based QE route. These routes generate two ranked lists of retrieved documents using the same sparse retriever, which are then fused using a modified reciprocal rank fusion method. Our empirical findings demonstrate that Exp4Fuse significantly improves the performance of both basic and advanced sparse retrieval models. We also provide a comprehensive discussion of the mechanisms behind Exp4Fuse, supported by extensive experimentation.

\section{Limitations }\label{sec:07}
Firstly, the Exp4Fuse framework leverages LLM-generated knowledge for QE, which is intrinsically linked to the quality of the underlying LLM. Identifying a suitable LLM is essential to fully demonstrate Exp4Fuse's capacity. Further investigations with various LLMs, such as GPT-4 and LLaMA 3, are warranted. Secondly, while Exp4Fuse theoretically conserves computational resources by using sparse retrieval and zero-shot prompting LLM, its reliance on LLM-generated hypothetical documents introduces certain latency. Therefore, future work should explore the computational efficiency of Exp4Fuse in greater detail.

\appendix
\section{Appendix}\label{A:01}

\subsection{Instructions}\label{A:0103}

\paragraph{TREC DL19} Instruction messages = \textit{"Please write a passage to answer the question. [question\_text]"}.

\paragraph{TREC DL20} Instruction messages = \textit{"Please write a passage to answer the question. [question\_text]"}.

\paragraph{MS MARCO dev} Instruction messages = \textit{"Please write a passage to answer the question. [question\_text]"}.

\paragraph{NQ} Instruction messages = \textit{"Please write a passage to answer the question. [question\_text]"}.

\paragraph{FiQA} Instruction messages = \textit{"Please write a financial article passage to answer the question. [question\_text]"}.

\paragraph{TREC\_NEWS} Instruction messages = \textit{"Please write a news passage about the topic. [question\_text]"}.

\paragraph{Robsut04} Instruction messages = \textit{"Please write a news passage about the topic. [question\_text]"}.

\paragraph{Touche2020} Instruction messages = \textit{"Please write a counter argument for the passage. [question\_text]"}.

\paragraph{DBPedia} Instruction messages = \textit{"Please write a passage to answer the question. [question\_text]"}.

\paragraph{SciFact} Instruction messages = \textit{"Please write a scientific paper passage to support/refute the claim. [question\_text]"}.

\subsection{Multiple query expansion}\label{A:0101}
In zero-shot LLM-based multiple query expansion, upon receiving an initial query $q_{o}$, a simple zero-shot prompt template — "\textit{Your task is to generate five different versions of the given question. [query]}" - generates five hypothetical queries, denoted as $Q = \{q_{1}, q_{2}, q_{3}, q_{4}, q_{5} \}$. These are then concatenated with the original query as input for the subsequent sparse retriever. To emphasize the original query in the augmented query, we implement a weight $\lambda_{1}$ that increases the length of the original query, with $\lambda_{1} = 2$ being an empirically effective value. Therefore, the augmented query $q_{me}$ generated by multiple query expansion with LLM can be formulated as follows:

\begin{equation}\label{03}
q_{me} = \textup{concat} (q_{o} \times \lambda_{1}, q_{1}, q_{2}, q_{3}, q_{4}, q_{5}).
\end{equation}

\subsection{Step-back query expansion}\label{A:0102}
In step-back prompting LLM-based query expansion, upon receiving a query $q_{o}$, a simple step-back prompt template — "\textit{What are the principles or mechanisms behind this question? [query]}" — is applied to generate a passage about the high-level concepts and first principles behind the query, denoted as $p_{p}$. This passage is then concatenated with the original query for input into the subsequent sparse retriever. To balance the influence of the original query and the generated passage, we implement a weight $\lambda_{2}$ that increases the length of the original query, with $\lambda_{2} = 5$ being an empirically effective value. Therefore, the augmented query $q_{sbe}$ generated by step-back prompting LLM can be formulated as follows:

\begin{equation}\label{04}
q_{sbe} = \textup{concat} (q_{o} \times \lambda_{2}, p_{p}).
\end{equation}

\end{document}